\documentclass{article}
\usepackage{frascatiphys,here,graphicx,subfigure}

\begin{document}

\title{ 
The jet vertex for Mueller-Navelet and forward jet production$^*$\\
 $_{\rm{^*Talk\; presented\; by\; Beatrice\; Murdaca.}}$ }
\author{
Francesco Caporale$^a$, Dmitry Yu.~Ivanov$^b$, 
Beatrice Murdaca$^{a}$, \\
Alessandro Papa$^a$, Amedeo Perri$^a$\\
{\em $^a$Universit\`a della Calabria and INFN, Gruppo collegato di Cosenza, 
Italy}\\
{\em $^b$Sobolev Institute of Mathematics and Novosibirsk State University, 
Russia}
}
\maketitle
\baselineskip=11.6pt
\begin{abstract}
We calculate in next-to-leading order BFKL the jet vertex relevant for the
production of Mueller-Navelet jets in proton collisions and of forward jets 
in deep inelastic scattering. Starting from the definition of the totally
inclusive quark and gluon impact factors in the BFKL approach and suitably 
considering the parton densities and the jet selection functions, we show that 
an infrared-free result can be found for the jet vertex. Finally we compare 
our expression for the vertex with the previous calculation of 
Refs.~\cite{bar1}.
\end{abstract}

\baselineskip=14pt

\section{Introduction}
\label{intro}

The Mueller-Navelet jet production process~\cite{Mueller:1986ey} was suggested
as an ideal tool to study the Regge limit of perturbative Quantum
ChromoDynamics (QCD) in proton-proton (or proton-antiproton) collisions.
The process under consideration is
\begin{equation}
p(p_1)+p(p_2)\to J_1(k_{J,1})+J_2(k_{J,2})+X \, ,
\label{process}
\end{equation}
where two hard jets $J_1$ and $J_2$ are produced.  Their transverse 
momenta are much larger than the QCD scale, $\vec k_{J,1}^2\sim \vec k_{J,2}^2
\gg \Lambda_{\rm QCD}^2$, so that we can use perturbative QCD. Moreover, they 
are separated by a large interval of rapidity, $\Delta y\gg 1$, which means 
large center of mass energy $\sqrt s$ of the proton collisions, 
$s=2p_1\cdot p_2 \gg \vec k_{J \, 1,2}^2$, since
$\Delta y\sim\ln {s/\vec k^2_{J\, 1,2}}$.
Since large logarithms of the energy compensate the small QCD coupling, they 
must be resummed to all orders of perturbative theory.
 
The BFKL approach~\cite{BFKL} is the most suitable framework for the
theoretical description of the high-energy limit of hard or semi-hard
processes. It provides indeed a systematic way to perform the resummation of 
the energy logarithms, both in the leading logarithmic approximation (LLA), 
which means resummation of all terms $(\alpha_s\ln(s))^n$, and in the
next-to-leading logarithmic approximation (NLA), which means resummation of
all terms $\alpha_s(\alpha_s\ln(s))^n$.

In QCD collinear factorization the cross section of the process reads
\begin{equation}
\frac{d\sigma}{dJ_1 dJ_2}
=\sum_{i,j=q,\bar q,g}\int\limits^1_0
\int\limits^1_0 dx_1dx_2 f_i(x_1,\mu) f_j(x_2,\mu)
\frac{d\hat \sigma_{i,j}(x_1 x_2 s,\mu)}{dJ_1 dJ_2}\;,
\label{ff}
\end{equation}
with $dJ_{1,2}=dx_{J_{1,2}}d^{D-2}k_{J_{1,2}}$  and the $i,j$ indices specify 
parton types (quarks $q$, antiquarks $\bar q$ or gluon $g$); $f_i(x,\mu)$ 
denotes the initial proton parton density function (PDF), the longitudinal
fractions of the partons involved in the hard subprocess are $x_{1,2}$,
$\mu$ is the factorization scale and $d\hat \sigma_{i,j}(x_1 x_2 s,\mu)$ is the
partonic cross section for the production of jets, $\hat s=x_1 x_2 s$ being
the energy of the parton-parton collision.
In the BFKL approach the resummed cross section of the hard subprocess is 
represented as the convolution of the jet impact factors of the colliding 
particles with the Green's function $G_\omega$, process-independent and 
determined through the BFKL equation,
\begin{equation}
\frac{d\hat \sigma}{dJ_1 dJ_2}=
\frac{1}{(2\pi)^{D-2}}\!\int\frac{d^{D-2}\vec q_1}
{\vec q_1^{\,\, 2}}\frac{d\Phi_{J,1}(\vec q_1,s_0)}{dJ_1}\!\int
\frac{d^{D-2}\vec q_2}{\vec q_2^{\,\,2}}\frac{d \Phi_{J,2}(-\vec q_2,s_0)}
{dJ_2}
\end{equation}
\vspace{-0.3cm}
$$\times\!\!\int\limits^{\delta +i\infty}_{\delta
-i\infty}\frac{d\omega}{2\pi i}\left(\frac{\hat s}{s_0}\right)^\omega
\!G_\omega (\vec q_1, \vec q_2)\, .
$$
The aim of this work is to illustrate the calculation of the NLA jet vertex. 

Since jets are originated by the hadronization of produced partons, the 
starting point is the impact 
factors for colliding partons~\cite{fading, fadinq, Cia, Ciafaloni:1998kx}. 
In order to select the partons that will generate the jet, we ``open'' one of 
the integrations over the partonic phase space and introduce a suitably 
defined jet selection function $S_J$. For the LLA impact factor, where there 
can be only a one-particle intermediate state, the jet function identifies 
the jet momentum with the momentum of the one parton ($S_J^{(2)}$). For the 
NLA impact factor we can have only either a one-particle  
(virtual corrections) or two-particle intermediate states. In the last case 
the $S_J$ function identifies the jet momentum with the momentum of one of 
the two partons or with the sum of the momenta of two partons ($S_J^{(3)}$).

In the calculation of the jet vertex, infrared divergences related with soft
emission will cancel in the sum with virtual corrections. The remaining 
infrared divergences are taken care of by the PDFs' renormalization.
The collinear counterterms appear due to the replacement of the bare PDFs by
the renormalized physical quantities obeying DGLAP evolution equations (in
the $\overline{\rm MS}$ factorization scheme). Ultraviolet divergences are 
removed by the counterterm related with QCD charge 
renormalization (in the $\overline{\rm MS}$ scheme).

Starting from the known lowest-order parton impact 
factors~\cite{fading,fadinq}, corresponding to the totally inclusive process,
we get the LLA jet impact factor by suitably introducing the $S_J^{(2)}$ 
function
\begin{equation}
\frac{d\Phi^{(0)}_J(\vec q\,)}{dJ}=\Phi^{(0)}_q \int_0^1 dx
\left(\frac{C_A}{C_F}f_g(x)+\sum_{a=q, \bar q} f_{a}(x)\right)S_J^{(2)}
(\vec q;x)\;,
\end{equation}
where $\Phi^{(0)}_q=g^2\frac{\sqrt{N_c^2-1}}{2N_c}$ is the quark impact 
factor (defined as the imaginary part of the quark-Reggeon diffusion process)
at the Born level and $\vec q$ is the Reggeon momentum.
\\
Substituting here the bare QCD coupling and bare PDFs by the renormalized ones,
we obtain the following expressions for the counterterms:
$$
\frac{d\Phi_J(\vec q\,)|_{\rm{charge \ c.t.}}}{dJ}
=   \frac{\alpha_s}{2\pi}\left(\frac{1}{\hat \varepsilon}+\ln\frac{\mu_R^2}
{\mu^2}\right)\left( \frac{11C_A}{6}-\frac{N_F}{3} \right)\, \Phi^{(0)}_q
$$
\vspace{-0.3cm}
\begin{equation}
\label{charge.count.t}
\times\int_0^1 dx\left(\frac{C_A}{C_F}f_g(x)+\sum_{a=q, \bar q} f_{a}(x)
\right) S_J^{(2)}(\vec q;x)
\label{ccharge}
\end{equation}
for the charge renormalization, and
\begin{equation}
\frac{d\Phi_J(\vec q\,)|_{\rm{collinear \ c.t.}}}{dJ}
= - \frac{\alpha_s}{2\pi}\left(\frac{1}{\hat \varepsilon}+\ln\frac{\mu_F^2}
{\mu^2}\right)\Phi^{(0)}_q
\int\limits^1_{0} \, d\beta \,  \int\limits^1_{0} \,dx\,
S_J^{(2)}(\vec q\,;\beta x)
\label{cpdf}
\end{equation}
\vspace{-0.3cm}
$$
\times\left[ \sum_{a=q,\bar q}\left( P_{qq}(\beta)f_{a}\left(x\right)
+  P_{qg}(\beta) f_g\left(x\right) \right)
\right.
$$
$$
\left.
+\frac{C_A}{C_F}\left( P_{gg}(\beta) f_g\left(x\right)
+  P_{gq}(\beta)\sum_{a=q,\bar q}f_{a}\left(x\right)\right)
\right]\;,
$$
for the collinear counterterm. 
\\
Now we have all the necessary ingredients to perform our calculation of the NLA
corrections to the jet impact factor. 

We will consider separately the subprocesses initiated by the quark and the
gluon PDFs and denote
\begin{equation}
V=V_q+V_g \, \ \ \ {\rm with} \ \ \  
\frac{d\Phi^{(1)}_J(\vec q\,)}
{dJ}\, \equiv \, \frac{\alpha_s}{2\pi}\,  \Phi^{(0)}_q \, V (\vec q\,)\; .
\end{equation}

\section{NLA jet impact factor}
\subsection{The quark contribution}

Virtual corrections are the same as in the case of the inclusive quark impact
factor~\cite{fading,fadinq,Cia}:
\begin{equation}
V_q^{(V)}(\vec q\,)=-\frac{\Gamma[1-\varepsilon]}{\varepsilon \,
(4\pi)^\varepsilon}\frac{\Gamma^2(1+\varepsilon)}{\Gamma(1+2\varepsilon)}
 \int_0^1 dx \sum_{a=q, \bar q}
f_{a}(x) \, S_J^{(2)}(\vec q\,;x)
\label{qvirt}
\end{equation}
\vspace{-0.3cm}
$$
 \times \left[C_F\left(\frac{2}{\varepsilon}-3\right)-\frac{N_F}{3}
+C_A\left(\ln\frac{s_0}{\vec q^{\,\,2}}+\frac{11}{6}\right)
 \right] + {\rm finite \ terms} \, .
$$

For the incoming quark case, real corrections originate from the quark-gluon
intermediate state. We denote the momentum of the gluon by $k$, then the
momentum of the quark is $q-k$; the longitudinal fraction of the gluon
momentum is denoted by $\beta x$. Thus, the real contribution has the 
form
$$
V^{(R)}_q\left( \vec q\right)=\int_0^1 dx\sum_{a=q, \bar q} f_{a}(x)
\left\lbrace\frac{\Gamma[1-\varepsilon]}{\varepsilon \,(4\pi)^\varepsilon}
\frac{\Gamma^2(1+\varepsilon)}{\Gamma(1+2\varepsilon)} 
\left[ C_F\left( \frac{2}{\varepsilon}-3\right)\,
S_J^{(2)}(\vec q\,;x)\,\right.\right.
$$
\vspace{-0.3cm}
\begin{equation}
\label{CF_quark}
\left.+ \int_0^1 d\beta \, \left(\,P_{qq}\left( \beta \right)
+\frac{C_A}{C_F}P_{gq}\left( \beta \right)\right)S_J^{(2)}\left( \vec q; x\beta\right)\right] 
+\frac{C_A}{(4\pi)^{\varepsilon}}\int \frac{d^{D-2}\vec k}
{\pi ^{1+\varepsilon}}
\label{qreal}
\end{equation}
\vspace{-0.3cm}
$$\left.
\times\frac{\vec q^{\,\, 2}}{\vec k^{\,\, 2}\left( \vec q-\vec k\right)^2}
\ln\frac{s_0}{ \left(|\vec k| + |\vec q-\vec k |\right)^2}\, S_J^{(2)}
\left( \vec q-\vec k; x\right)\right\rbrace+ {\rm finite \ terms}\ .
$$

\subsection{The gluon contribution}

Virtual corrections are the same as in the case of the inclusive gluon impact
factor~\cite{fading,Cia}:
\begin{equation}
V_g^{(V)}(\vec q)=-
\frac{\Gamma[1-\varepsilon]}{\varepsilon\,  (4\pi)^\varepsilon}
\frac{\Gamma^2(1+\varepsilon)}{\Gamma(1+2\varepsilon)}\,
\int_0^1 dx \, \frac{C_A}{C_F}\, f_g(x) \, S_J^{(2)}(\vec q\,;x) 
\label{gvirt}
\end{equation}
\vspace{-0.3cm}
$$
\times\left[\, C_A\ln\left(\frac{s_0}{\vec q^{\: 2}}\right)
+C_A\left( \frac{2}{\varepsilon}-\frac{11}{6}\right) + \frac{N_F}{3} \right]
+ {\rm finite \ terms} \; .
$$

In the NLA gluon impact factor real corrections come from intermediate states
of two particles, which can be quark-antiquark or 
gluon-gluon~\cite{fading,Cia,Ciafaloni:1998kx}.

We find
$$
V_g^{(R)}(\vec q\,)=  \frac{\Gamma[1-\varepsilon]}{\varepsilon\,  
(4\pi)^\varepsilon} \frac{\Gamma^2(1+\varepsilon)}{\Gamma(1+2\varepsilon)}
\,\int_0^1 dx  \, f_g(x) \, \left\lbrace \frac{C_A}{C_F}\left(\frac{N_F}{3}
+\frac{2C_A}{\varepsilon}-\frac{11}{6}C_A\right)\,\right.
$$
$$
\times S_J^{(2)}(\vec q;x)+\,
\int_0^1 d\beta \, \left[2N_F P_{qg}(\beta)+2C_A\frac{C_A}{C_F}\left(P(\beta)
+\frac{(1-\beta)P(1-\beta)}{(1-\beta)_+}\right)\right]\,
$$
$$
\left.\times S_J^{(2)}(\vec q;x\beta) \right\rbrace
+\, \frac{C_A }
{(4\pi)^{\varepsilon}} \int_0^1 dx \, \frac{C_A}{C_F} \, f_g(x)
\int \frac{d^{D-2}\vec k}{\pi^{1+\varepsilon}}
\frac{\vec q^{\: 2}}{\vec k^{ 2}(\vec k-\vec q\,)^2}
\ln \frac{s_0}{(|\vec k|+|\vec q-\vec k|)^2} 
$$
\vspace{-0.3cm}
\begin{equation}
\times\, S^{(2)}_J(\vec q-\vec k;x)+\,  {\rm finite \ terms}\; .
\label{greal}
\end{equation}

To conclude, we collect the contributions given in 
Eqs.~(\ref{ccharge}),~(\ref{cpdf}),~(\ref{qvirt}), 
~(\ref{qreal}),~(\ref{gvirt}),~(\ref{greal}), and we note that we are left 
with two divergences: the last 
of~(\ref{qreal}) and of~(\ref{greal}). It easy to see that the integration of 
those terms over $\vec q$ with any function, regular at $\vec q=\vec k_J$, 
will give a divergence-free result. In particular, a finite result will be 
obtained 
after the convolution of the jet vertex with BFKL Green's function, which is
required for the calculation of the jet cross section.
\\
Note that divergent terms of the  two parton intermediate state 
contributions, shown in Eqs.~(\ref{qreal}) and~(\ref{greal}),
are expressed through the jet function $S_J^{(2)}$, due to reduction of 
$S_J^{(3)}\to S_J^{(2)}$ in the kinematic regions of soft or collinear parton 
radiation.
\\
More details about this calculation can be found in Ref.~\cite{Caporale}.

\section{Summary}

We have recalculated the jet vertices for the cases of quark and gluon in
the initial state, first found by Bartels {\it et al.}~\cite{bar1}. Our 
approach is different, since the starting point of our calculation is the 
known general expression for next-to-leading-order impact factors, given in 
Ref.~\cite{FF98}, applied to the special case of partons in the initial state. 
Nevertheless, in many technical steps we followed closely the derivation of 
Refs.~\cite{bar1}.
\\
In our approach the energy scale $s_0$ remains untouched and need not
be fixed at any definite value. In order to compare our results with those
of~\cite{bar1}, we need to perform the transition 
(see~\cite{Fadin:1998sg}) from the standard BFKL scheme with arbitrary energy
scale $s_0$ to the one used in~\cite{bar1}, where the energy scale
depends on the Reggeon momentum. After this procedure, we can see a complete 
agreement with~\cite{bar1}.
\\
The jet vertex discussed in this paper is an essential ingredient also
for the study of the inclusive forward jet production in deep inelastic 
scattering in the NLA.

\end{document}